\documentclass[doublecol]{epl2}

\usepackage{dsfont}

\title{Simultaneous spin-based Boolean logic operations with re-programmable
functionality}

\shorttitle{Simultaneous Boolean logical operations}

\author{Moumita Patra\inst{1} \and Santanu K. Maiti\inst{1}\thanks{E-mail:
\email{santanu.maiti@isical.ac.in}}}

\shortauthor{Moumita Patra \em et al.}

\institute{
 \inst{1} Physics and Applied Mathematics Unit, Indian Statistical
Institute, 203 Barrackpore Trunk Road, Kolkata-700 108, India
}

 {Spin polarized transport}
 {Reprogrammable magnetic logic}

\abstract{Construction of parallel logic gates at nano-scale level 
undoubtedly improves the efficiency of computable operations. In this work 
we put forward a new idea of designing two distinct logical operations 
{\em simultaneously} in the two output leads of a three-terminal bridge 
setup which on one hand is highly stable as all the results are valid 
for a wide range of parameter values and on the other hand easy to engineer.
Our system can be reprogrammed to have all the two-input logic gates with
two operations at a time by selectively choosing the physical parameters 
describing the system, viz, Rashba spin-orbit (SO) interaction, magnetic 
flux, and Fermi energy. Finally, we explore the possible storage mechanism 
as well using our model.}

\begin{document}

\maketitle

\section{Introduction}

The logic gates are the most essential bricks of modern computers and 
digital electronics as their functionalities rely on the implementation
of Boolean functions. These gates are usually comprised of various
field-effect transistors (FETs) and metal oxide semiconductor field
effect transistors (MOSFETs). So, finding of logical responses in a
simple nano-scale device is a subject of intense research for better
performance of computable operations. And for proper execution of such 
operations, wiring between individual logic gates is definitely required 
which limits integration densities, gives rise to huge power consumption 
and restricts processing speeds~\cite{cite1}. Therefore, accommodation 
of Boolean logic gates into a single active element is highly desirable 
to eliminate wiring amongst transistors.

Though a wealth of literature knowledge has been developed in designing 
logic gates essentially based on molecular systems~\cite{cite2,cite3,cite4}, 
but most of these works are involved in single logic operation at a time, 
and very less number of works are available so far in the context of 
functioning parallel logic gates in one setup which is highly desirable 
from the efficiency perspective and suitable computable operations. 
Hod {\em et al.}~\cite{hod} have made an effort to design parallel logic 
gates considering a cyclic molecule where a realistic magnetic field and 
gate potential are used as the inputs. In their work they have only shown 
AND and NAND operations. A completely different prescription was given by
imposing a novel architecture considering a single parametric resonator
(electromechanical) where three logical operations along with multibit 
logic functions can be performed~\cite{cite1}. This work essentially 
suggests a suitable prospect of designing parallel logic processor using 
a single resonator. There are other few realizations of parallel logic 
operations~\cite{pl1,pl2,pl3,pl4} considering different semi-conducting 
materials, molecular systems, protein-like molecules, synthetic gene 
networks. But these studies do not essentially address the phenomenon of 
`simultaneous Boolean logical operations', which is precisely our main 
motivation of the present work.

Most of the works available in literature exploit electronic charge for
logical operations, but the implementation of these functional operations
based on spin degree of freedom undoubtedly yields several advantages like
rapid processing, much smaller energy consumptions, greater integration 
densities, etc.~\cite{spin1,spin2}. In order to design an efficient 
spintronic device be it logic functions or any other operations, we need 
to take care about two important things: spin injection efficiency and 
spin coherence length. Metallic systems are much superior than semiconducting 
materials in the aspect of spin injection, but the previous ones have much 
lower spin coherence length~\cite{metal}. 
Both these two facts viz, efficient spin injection and coherence length,
can be incorporated if we can construct the device using normal metal
by compromising on system size. Hopefully it can be done with suitable 
designing of the setup, and we explore it in this article. Here we also
circumvent the consideration of molecular systems, as normally used in
describing logic operations, due to the fact that they exhibit much lower
transconductance~\cite{cite5}. 

\begin{figure}[ht]
{\centering \resizebox*{5cm}{3.5cm}{\includegraphics{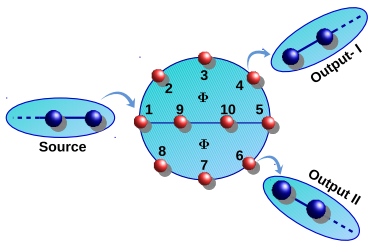}}\par}
\caption{(Color online) Schematic representation of three-terminal setup 
where two logical operations are obtained simultaneously at two out going 
leads. Up and down spin electrons are injected from the source lead. These 
leads are coupled to the ring which plays the central role of all logical 
operations. The full ring is equally divided into two parts, and based on 
the requirement of logical operations, in some cases, we apply a magnetic 
flux $\Phi$. The ring circumference is subjected to DSOI and RSOI, whereas 
the central horizontal line is free from any kind of SO interaction.}
\label{fig1}
\end{figure}

Considering all these factors, here we propose a new idea of designing 
`simultaneous Boolean logic operations' using a three-terminal bridge
setup (see Fig.~\ref{fig1}) where the output response is fully spin based. 
In the two outgoing leads we get two different logical operations at an
identical time which we measure by calculating spin current $I_s$, and
the central mechanism is controlled by the system placed within the three
contacting leads. The bridging system consists of a metallic ring which is
divided equally to form two sub-rings. Apart from the common portion of the 
two sub-rings (viz, the dividing line connecting the sites $1$, $9$, $10$ 
and $5$), the rest section (i.e., the ring circumference joining the sites 
$1$, $2$, $\dots$, $7$ and $8$) is subjected to Dresselhaus SO 
interaction (DSOI)~\cite{dsoi}, and it is distributed uniformly throughout 
the ring. Along with this we also consider Rashba SO interaction 
(RSOI)~\cite{rsoi} where two different cases, viz, uniform and non-uniform, 
are considered for the distribution of RSOI along the ring circumference to
implement specific simultaneous logic operations, and it will be clearly 
observed from our subsequent analysis. Both these two SO interactions are 
commonly encountered in solid state materials and 
among them RSOI draws much attention as its strength can be tuned 
externally~\cite{gate1,gate2} which yields controlled spin transmission. 
We use RSOI as one of the input signals of logic operations, and, in some 
cases we also introduce equal amount of magnetic flux in the two sub-rings
which is treated as another input signal. The `ON' and `OFF' states of the 
output signal are described by the positive and negative signs of $I_s$, 
respectively, where $I_s=I_{\uparrow}-I_{\downarrow}$ 
($I_{\sigma (\sigma=\uparrow,\downarrow)}$ being the spin dependent current). 
By selectively choosing the physical parameters, viz, RSOI, magnetic flux and
Fermi energy, the present setup can be `reprogrammed' to have all the six 
two-input Boolean logic gates with two operations at a time. Achieving these 
parallel logic operations we can also think about other special-purpose logic 
operations~\cite{repro} like full-adder, half-adder, multiplier, switching 
spin action, etc.

The rest of the paper is arranged as follows. With the above brief 
introduction and motivation, next we illustrate our model quantum system and 
the theoretical prescription for the calculations. The logical operations 
are clearly described in a separate section. In this section we also 
discuss the possibilities of utilizing the setup for storage mechanism.
Logical operations along with storage function is extremely important for the
complete executation of computable operation and that is hopefully possible as
our response is spin based. In usual charge based devices we need to transfer
the information to a memory as these are usually highly volatile~\cite{repro}.
Finally, we end with conclusion and future perspectives of spintronic 
applications.

\section{Model Hamiltonian and the Method}

The full bridge system described in Fig.~\ref{fig1} is divided into three parts:
the central ring conductor, three leads (one incoming and two outgoing), and 
conductor-to-lead coupling. We simulate these parts by the tight-binding (TB) 
framework. Assuming the leads are perfect and semi-infinite, we can write the
TB Hamiltonian of the leads as
\begin{equation}
H_{\mbox{\tiny leads}} = \sum\limits_p \Big[ \sum\limits_i
c_i^\dagger \epsilon_0 \mathds{1} c_i + \sum\limits_i \left(
c_{i+1}^\dagger t_0 \mathds{1} c_n +  h.c.\right)\Big]
\label{eq1}
\end{equation}
where the summation over $p$ ($p$ runs from $1$ to $3$) is used for the three 
leads. The parameters $\epsilon_0$ and $t_0$ describe the on-site energy and 
nearest-neighbor hopping (NNH) integral, respectively. For an ordered lead we 
can easily put $\epsilon_0=0$, without loss of any generality. $t_0$ controls 
the band width ($4t_0$) of the leads. We couple the incoming lead at site $1$ 
of the ring, for the entire analysis, whereas the other two leads are connected
at two other sites of the ring (say, $k$ and $l$), those are variables. 
The leads are coupled to the ring through the hopping parameter $\tau_p$.

The TB Hamiltonian for the central system looks quite different from 
Eq.~\ref{eq1}, as the ring system is subjected to Rashba and Dresselhaus 
SO interactions, and the magnetic flux as well. The dividing line is free from 
any kind of SO interaction, and since the two sub-rings are threaded by equal
amount of magnetic flux, no phase factor will introduce into this segment.  
We write the general Hamiltonian of the central ring (CR) geometry 
as~\cite{ham1,ham2,ham3,ham4}
\begin{eqnarray}
H_{\mbox{\tiny CR}}
&=& \sum\limits_{n (\mbox{\tiny all sites})} c_n^\dagger
\epsilon_n \mathds{1} c_n + \sum\limits_{n (\mbox{\tiny wire})}
\left(c_{n+1}^\dagger t \mathds{1} c_n + h.c.\right) \nonumber \\
& + &\sum\limits_{n (\mbox{\tiny ring})}\Big[c_{n+1}^\dagger
t_D e^{i\theta}(\mbox{$\sigma_y$}\cos\zeta_{n,n+1} 
+\mbox{$\sigma_x$}\sin\zeta_{n,n+1})c_n\nonumber \\
&+& h.c. \Big] - i\sum\limits_{n (\mbox{\tiny ring})}
\Big[c_{n+1}^\dagger t_R e^{i\theta}(\mbox{$\sigma_x$}\cos
\zeta_{n,n+1}+ \nonumber \\
& & \mbox{$\sigma_y$} \sin \zeta_{n,n+1}) + h.c.\Big]
\label{eq2}
\end{eqnarray}
where $c_n$ is a column of operators formed with the fermionic operators
$c_{n\uparrow}$ and $c_{n\downarrow}$. $\theta=\pi \Phi/2$ is the phase 
factor acquired by an electron~\cite{ph} while traversing through the 
periphery of the ring. In this Hamiltonian we do not consider any spin
splitting mechanism due to Zeeman interaction, as it is too small compared
to the other two splitting mechanisms associated with the Rashba and 
Dresselhaus SO couplings. With this assumption no physical picture will 
be altered. 
The Rashba and Dresselhaus SO interactions are described by the factors 
$t_R$ and $t_D$, respectively, and $\zeta_n=2\pi(n-1)/N$ ($N$ being the 
total number of atomic sites in the ring, and for our schematic diagram it 
is $8$) which defines the factor $\zeta_{n,n+1}=(\zeta_n+\zeta_{n+1})/2$. The
other physical parameters $\epsilon_n$ and $t_n$ represent on-site energy and 
NNH integral in the ring as well as in the central wire. 
$\sigma_i$'s ($i=x,y,z$) are the usual Pauli spin matrices where $\sigma_z$ 
is diagonal.

This is all about the model and the TB Hamiltonians describing the full 
system. Now, in order to describe the logical responses in two outgoing 
leads we need to calculate spin currents. At absolute zero temperature, 
the spin current at $q$th lead ($q$ can be lead-2 (i.e., output-I) and 
lead-3 (i.e., output-II)) is computed from the relation~\cite{datta}
\begin{equation}
I_s^q(V) = \frac{e}{h}
\int\limits_{E_F-\frac{eV}{2}}^{E_F+ \frac{eV}{2}}T_{1q}(E) \, dE
\label{eq3}
\end{equation}
where $T_{1q}(E)$ is the effective two-terminal spin selective transmission 
probability, and it is defined as 
$T_{1q}(E)=(T_{1q}^{\uparrow\uparrow}+T_{1q}^{\downarrow\uparrow})-
(T_{1q}^{\downarrow\downarrow}+T_{1q}^{\uparrow\downarrow})$. To find spin 
dependent transmission probabilities $T_{1q}^{\sigma\sigma^{\prime}}$
we use Green's function method, and in terms of the retarded and advanced 
Green's functions ($G^r$, $G^a$) it can be expressed as~\cite{datta,car,fl} 
$T_{1q}^{\sigma\sigma^{\prime}}(E) = \mbox{Tr}\big[\Gamma_1^{\sigma} 
G^r \Gamma_q^{\sigma^{\prime}} G^a\big]$, where 
$\Gamma_1^{\sigma}, \Gamma_q^{\sigma^{\prime}}$ are the coupling matrices and 
$G^r=(G^a)^{\dagger}=(E-H_{eff})^{-1}$. $H_{eff}$ is the effective Hamiltonian 
of the central ring system by incorporating the effects of side-attached leads 
through self-energy corrections. In our prescription, positive $I_s$ means 
high output, while negative $I_s$ corresponds to the low output.

\section{Essential Results and Discussion}

{\em Simultaneous logical operations:} As already stated above, by selectively 
choosing physical parameters, viz, Rashba SO coupling, magnetic flux $\Phi$
in each sub-rings and location of the outgoing leads we can design all possible 
Boolean logic gates, two such gates at a time. Here we present three pairs 
(OR-NOR, AND-NAND, and XOR-XNOR) for a specific set of parameter values, as
illustrative examples, but one can get other different pairs quite easily 
simply by adjusting the required variables, and thus, our system is 
reprogrammable.

We carry out numerical calculations at absolute zero temperature, considering
a $10$-site system as discussed in Fig.~\ref{fig1}. Throughout the analysis we 
set, unless otherwise specified, all site energies to zero, NNH integral in 
contacting leads at $2\,$eV, and the rest other NNH integrals including 
ring-to-lead coupling at $1\,$eV. The DSOI is fixed at $0.25\,$eV, and it is
uniform throughout the ring circumference as stated earlier. The other
two physical parameters, RSOI and $\Phi$, are no longer constant and we 
mention their specific values during the subsequent analysis. In what 
follows we present different functional logical operations one by one.

\begin{figure}[ht]
{\centering \resizebox*{8cm}{6cm}{\includegraphics{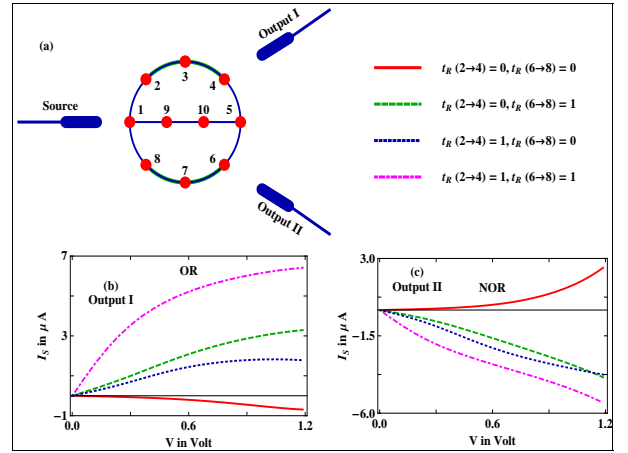}}\par}
\caption{(Color online) (a) Cartoon of the setup where two different values of
RSOI are considered as two input states and these are introduced in the two 
green portions of the upper and lower arms of the ring. In the rest blue 
portions of the ring circumference the RSOI strength is fixed at $0.5\,$eV.
In the spectra (b) and (c), the simultaneous logical operations are shown by 
plotting spin currents, at two outgoing leads, as a function of bias voltage.
Here we choose $E_F=0.4\,$eV.}
\label{fig2}
\end{figure}

\vskip 0.25cm
\noindent
{\underline{Case I. OR and NOR operations:}} The setup is shown in 
Fig.~\ref{fig2}(a) where the outgoing leads are coupled to sites $4$ and $6$,
respectively, of the ring. Here two different strengths of RSOI are taken into 
account those are treated as low and high states of the inputs, and no magnetic 
flux is added. These two input states are implemented by changing the Rashba 
strengths at the green portions of the upper and lower arms of the ring 
(see Fig.~\ref{fig2}(a)), keeping a constant magnitude of RSOI in the other 
parts. It looks like a hybrid ring and seems quite easy to fabricate. 
The responses for this setup at the two outgoing leads under different 
input conditions are shown in Figs.~\ref{fig2}(b) and (c). The spin current 
$I_s$ is computed up to a reasonable bias voltage, and for this entire 
voltage window we can clearly see that the two outgoing leads exhibit two 
different logical operations (OR and NOR) simultaneously. The output 
currents are also sufficiently high ($\sim \mu$A) which thus easy to detect. 
The underlying physics involved relies on the interplay between RSOI 
and DSOI which leads to anisotropic spin dependent transport in the outgoing
leads as discussed clearly by Chang and co-workers~\cite{ham1,chnew,chnew1}. 
In presence of both the two SO couplings, an effective periodic potential 
is developed which breaks the rotational symmetry of the ring, resulting 
non-trivial spin dependent transport phenomena~\cite{ham1,chnew,chnew1}. 
To achieve simultaneous logical operations we essentially need to get 
polarized spin currents from an unpolarized beam of electrons in outgoing 
leads of a multi-terminal bridge setup. Several propositions have already 
been made by some groups and by one of the authors of us along this direction 
\begin{figure}[ht]
{\centering \resizebox*{8cm}{5.5cm}{\includegraphics{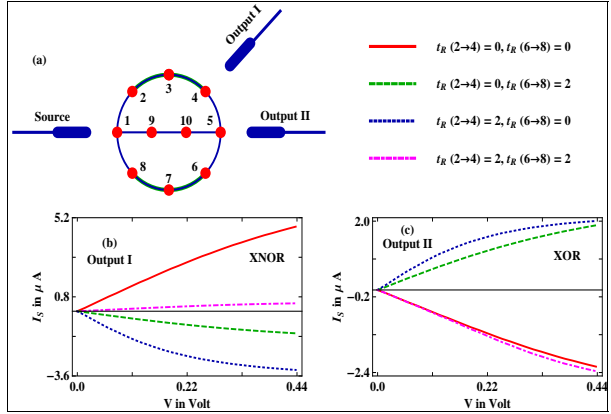}}\par}
\caption{(Color online) Junction setup (a) together with simultaneous logical
operations (XOR and XNOR) ((b) and (c)) for low ($t_R=0$) and high ($t_R=2\,$eV)
input states. The RSOI strength at the blue portion of the ring circumference is
fixed at $0.25\,$eV, and magnetic flux is not required like the previous case 
(viz, Fig.~\ref{fig2}). The currents are calculated by considering $E_F=0.4\,$eV.}
\label{fig3}
\end{figure}
i.e., how to get polarized spin currents in presence of SO interactions in 
outgoing leads considering different shaped 
geometries~\cite{spl1,spl2,spl3,spl4,spl5}. The main focus of those works 
was to achieve polarized spin currents under different input conditions, 
but no one has attempted to think about logical operations, {\em especially 
simultaneous logic functions.} This is precisely what we do in our present 
work, and the responses what we get in two outgoing leads are basically the 
combined effect of SO interactions and quantum interference of electronic 
waves passing through different sectors of the geometry. 
Here it is important to note that all the logical operations are 
implemented by determining the spin current $I_s$, and more precisely by 
noting its sign viz, positive or negative. Thus, for two logical operations at
the two output leads, we need to satisfy all the operations simultaneously
(a set of four outputs for each logic gates) associated with the input 
conditions, and we achieve this goal considering the interplay between the 
RSOI and DSOI, and the interference among the electronic waves. If we set 
any one the two SO interactions to zero, which brings back the rotational 
symmetry in the ring~\cite{spl1,spl2,spl3,spl4}, it will be too hard 
to satisfy all the above mentioned operations at the two output leads. 
Particularly, when DSOI becomes zero (for instance), no spin current will be
available for the input condition where RSOI is also zero, which thus
fails to explain logic functions. In that case we have to consider non-zero 
Rashba couplings for the inputs, but satisfying all the output conditions 
will not be quite simple unlike the cases we discuss here with our present 
setups. 

\vskip 0.25cm
\noindent
{\underline{Case II. XOR and XNOR operations:}} Considering the identical
ring type (viz, the hybrid ring where RSOI is distributed non-uniformly) as 
taken in Case I, and slightly modifying the location of one of the two outgoing 
\begin{figure}[ht]
{\centering \resizebox*{8cm}{6cm}{\includegraphics{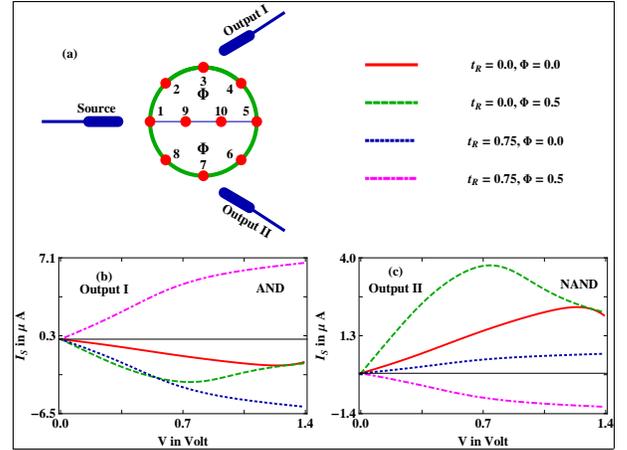}}\par}
\caption{(Color online) Setup for the AND and NAND operations in two outgoing
leads where RSOI is uniformly distributed along the ring circumference, unlike
the previous two configurations. This RSOI acts as one of the two inputs, while
for the other input we apply equal amount of magnetic flux $\Phi$ in each
sub-rings. The results are shown in (b) and (c), where these currents are 
computed for $E_F=-1.35\,$eV. Tuning of Fermi energy can be substantiated 
with suitable gate electrodes~\cite{gate3,gate4}.}
\label{fig4}
\end{figure}
leads we get a pair of another two simultaneous logical operations. The setup 
along with the results are placed in Fig.~\ref{fig3}, where we see that XOR and 
XNOR operations are clearly obtained from the two outgoing leads. We simulate 
these results setting the equilibrium Fermi energy at $0.4\,$eV. Comparing the
results given in Figs.~\ref{fig2} and \ref{fig3} we get a clear hint about the
robust effect of quantum interference as in one case a specific set of two 
logical operations are obtained, while another such set is visible for the 
other case.
\begin{figure*}[ht]
{\centering
\resizebox*{8cm}{6cm}{\includegraphics{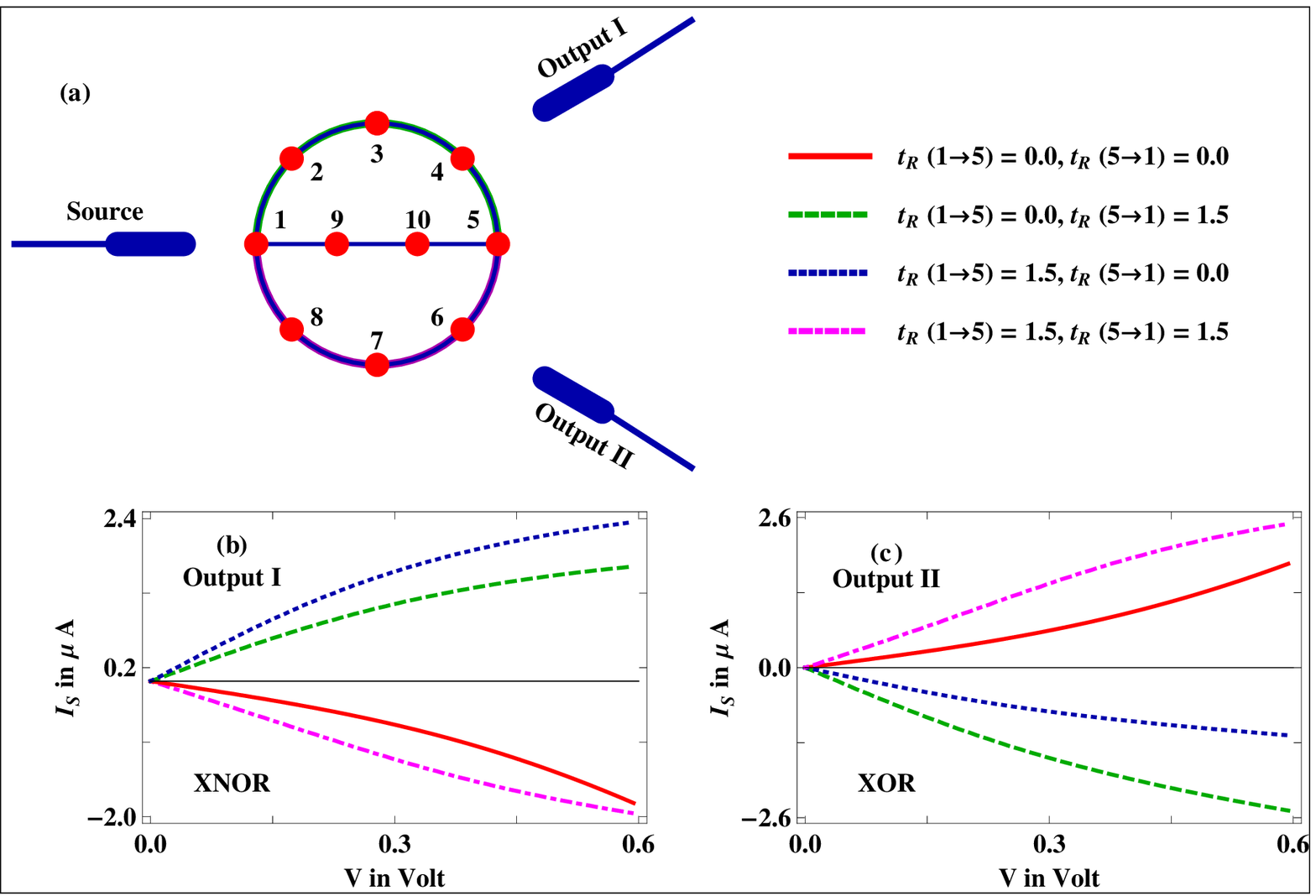}}
\resizebox*{8cm}{6cm}{\includegraphics{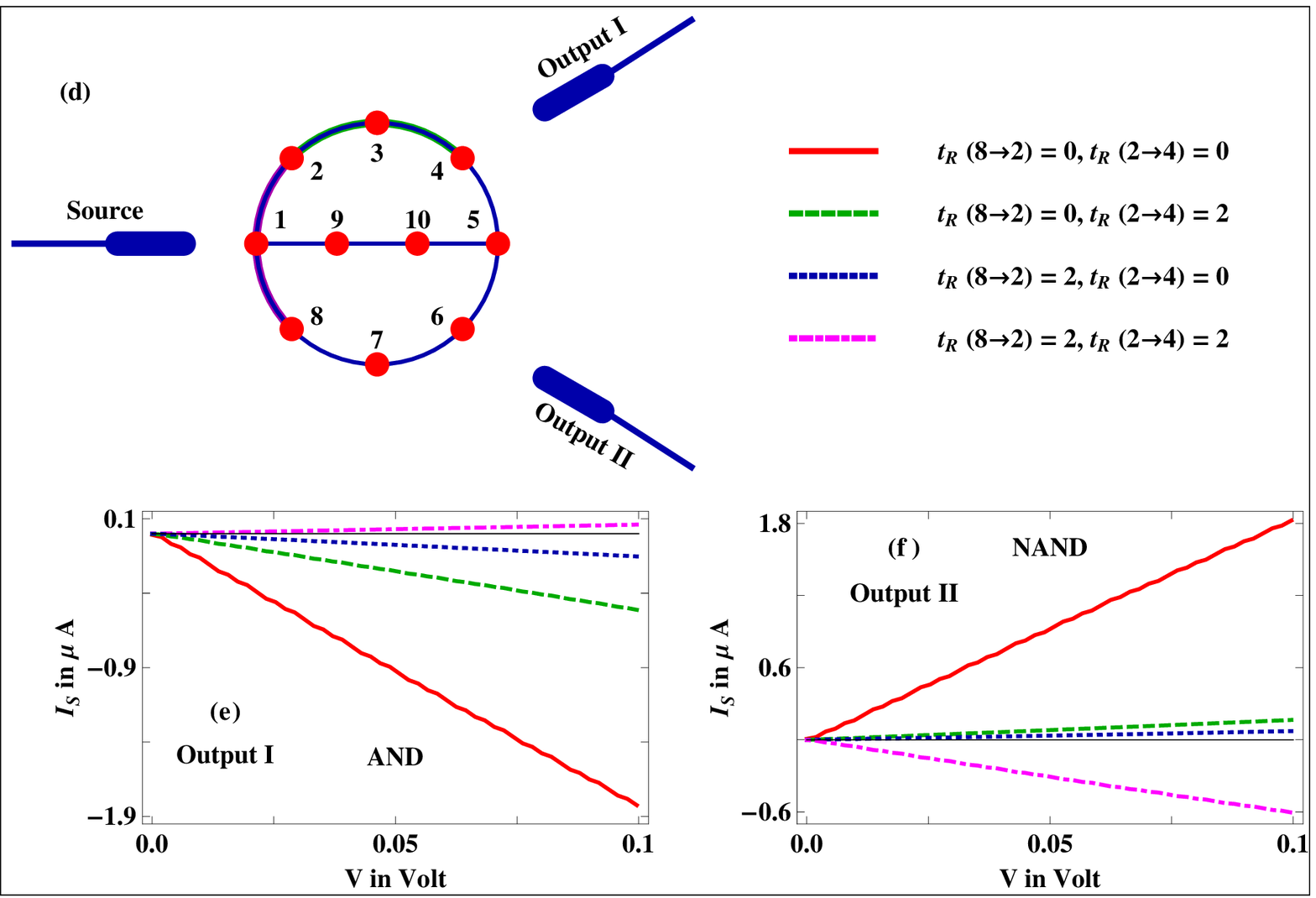}}
\par}
\caption{(Color online) Implementation of all possible logic operations 
form a single ring-lead geometry. In this figure four logic functions 
(XOR-XNOR and AND-NAND) are shown, while for this same configuration 
other two operations (OR-NOR) are presented in Fig.~\ref{fig2}. No 
magnetic flux is included
here and the DSOI strength is kept unchanged as before. The inputs are given 
in terms of different Rashba SO couplings, those are introduced in the thick 
green and magenta regions. The rest blue portions of the setups are free 
from any RSOI. For the XOR-XNOR operations we choose $E_F=0.3\,$eV, while 
it is fixed at $-0.25\,$eV for the AND-NAND functions.}
\label{fig5}
\end{figure*}

\vskip 0.25cm
\noindent
{\underline{Case III. AND and NAND operations:}} Finally, we consider another
configuration to implement other two logic functions i.e., AND and NAND 
operations. Here the full circumference is subjected to Rashba SO interaction
which acts as one of the two input signals, and for the other input we impose 
equal amount of magnetic flux $\Phi$ in each of the two sub-rings as shown
schematically in Fig.~\ref{fig4}(a). Thus, RSOI and $\Phi$ are used for the 
two inputs of logic functions, and the responses in the two outgoing leads,
associated with four input conditions are placed in Figs.~\ref{fig4}(b) and (c).
The two logical operations (AND and NAND) are clearly visible, and in this case
the interplay between SO couplings, magnetic flux and quantum interference plays
the central role to exhibit these two logic operations.

\vskip 0.25cm
\noindent
{\underline{Case IV. Single setup - all possible logic operations:}}
To put more emphasis on reprogrammability, finally we search for a 
possible ring-lead configuration where all the two-input logic gates can 
be achieved. In Fig.~\ref{fig5} four logical operations (XOR-XNOR 
and AND-NAND) are presented for a specific ring-to-lead geometry, and 
interestingly, for this same setup other two logic functions (OR-NOR) 
are also implemented as discussed earlier in Fig.~\ref{fig2}. 
Looking carefully into the spectra and comparing the results given 
in Figs.~\ref{fig3}-\ref{fig5} one can see that the responses
obtained in Fig.~\ref{fig5} are quite inferior, as the magnitudes 
of spin currents in outgoing leads are reasonably lower in few cases, 
rather than that the individual geometries, i.e., the responses obtained in
Figs.~\ref{fig3} and \ref{fig4}. This low-current response hopefully
be sacrificed as we can able to establish all the possible two-input logic 
gates, two operations at a time, from a single ring-lead configuration. 
Thus, a possible hint of designing reprogrammable logic gates is expected.
This argument i.e., the reprogrammability can be strengthened further 
following the propositions given by Peeters and his group in a work where 
they have shown that programmable spintronic devices can be designed using 
a network of quantum rings in which selective spin transmission will be 
obtained by locally tuning the Rashba SO coupling in different rings of the 
network~\cite{qnano}.

Before we end the discussion of simultaneous logical operations, we would
like to note that one may ask whether the same functionality persists if we
consider a similar kind of geometry by removing the atomic sites $9$ and 
$10$ i.e., in the absence of the central horizontal line. The answer should
not be strictly no, but it is very difficult to execute all the six logic
functions, especially, two logic operations at the two outgoing leads which
we confirm through our detailed numerical calculations. It is true that 
the polarizing 
effect in presence of RSOI and DSOI, based on which the logic operations 
are designed, is available even in a single ring geometry with one input 
and two outgoing leads, but the inclusion of multiple paths to form a 
network always yields novel spintronic features, which is substantiated 
clearly in Refs.~\cite{qnano,qnano1}.

\vskip 0.25cm
\noindent
{\em Applicability as a storage device:} Along with the above mentioned 
functional logical operations here we give a brief outline how such a system 
can be utilized for storage purposes as well. Utilization of spin orientation 
($\uparrow$, $\downarrow$) for storing information will be the most suitable 
operation~\cite{spin1} as it does not alter its state unless some perturbations 
are imposed. The idea originated from the mechanism of spin-transfer torque 
(STT)~\cite{stt1} which suggests that a beam of polarized spin current having 
sufficient magnitude can rotate the spin orientation of a free magnetic moment, 
by transferring spin angular momentum, along the spin direction of the incident 
beam. Much higher spin current above cutoff for switching spin magnetization 
can easily be achieved~\cite{stt1} in our case mainly because of too 
narrow outgoing channel. Depending on the sign ($+$ve or $-$ve) of the 
polarized spin current $I_s$, the free magnetic moment aligns along $+Z$ or 
$-Z$ direction, and assigning 1 or 0 of the logic bits with these orientations 
we can eventually store one bit memory~\cite{stt1,stt2,stt3,stt4}. The free 
magnetic site can directly be embedded in the outgoing lead wire or be placed 
in its close proximity, and in either of these two cases angular momentum 
transfer takes place through exchange mechanism. Thus, for the present setup 
as there are two outgoing leads, we can think about two such free magnetic 
sites, and in principle, can store two bits simultaneously which significantly 
enhances the storage capacity.

\section{Closing Remarks}

In this work we make an in-depth analysis of designing simultaneous logic
gates based on spin states that has not been discussed so far in literature,
to the best of our concern. The significance of this proposal is that it
relies on a simple tailor made geometry that can be configured to achieve 
different functional logical operations. Though the magnitude of spin 
current $I_s$ slightly change with the strengths of SO fields and magnetic 
flux, all the essential results, determined by the sign of $I_s$, remain 
unchanged for a wide range of parameter values including bias voltage 
that we confirm through our exhaustive numerical calculations. Along 
with the logical operations, we also put forward an idea of devising 
this system for storage purposes utilizing the concept of spin exchange 
interaction. Since in this three-terminal setup, polarized spin currents 
are obtained at the two outgoing leads, we can in principle store two 
bits by imposing two free magnetic sites, which yields higher storage 
capacity. Thus, both logic functions and storage mechanism can be 
implemented in a single device, circumventing the use of additional storage
device as usually considered in charge based systems, which no doubt brings 
significant impact to hit the present market of nanotechnology and 
nanoengineering. Finally, we end our discussion by pointing out that this 
proposal of simultaneous Boolean logic operations can be generalized to have
more complex parallel logic operations by adding more output leads and 
re-programmed the system by the external factors.

\acknowledgments

First author (MP) would like to acknowledge the financial support of University 
Grants Commission, India (F. $2-10/2012$(SA-I)) for pursuing her doctoral work.


\begin{thebibliography}{0}

\bibitem{cite1} I. Mahboob, E. Flurin, K. Nishiguchi, A. Fujiwara,
and H. Yamaguchi,  Nat. Commun. \textbf{2}, 198 (2011).

\bibitem{cite2} A. P. de Silva, H. Q. N. Gunaratne, and C. P. McCoy,
Nature \textbf{364}, 42 (1993).

\bibitem{cite3} F. M. Raymo, Adv. Mater. \textbf{14}, 401 (2002).

\bibitem{cite4} A. P. de Silva {\it et al.} Chem. Rev. \textbf{97},
1515 (1997).

\bibitem{hod} O. Hod, R. Baer, and E. Rabani, J. AM. CHEM. SOC.
\textbf{127 (6)}, 1648 (2005).

\bibitem{pl1} B. Fresch, M. Cipolloni, T.-M. Yan, E. Collini, R. D.
Levine, and F. Remacle, J. Phys. Chem. Lett. \textbf{6}, 1714 (2015).

\bibitem{pl2} Y. Xu, X. Jin, and H. Zhang, Phys. Rev. E \textbf{88},
052721 (2013).

\bibitem{pl3} A. Dari, B. Kia, A. R. Bulsara, and W. L. Ditto, Europhys.
Lett. \textbf{93}, 18001 (2011).

\bibitem{pl4} H. Ando, S. Sinha, R. Storni, and K. Aihara, Europhys.
Lett. \textbf{93}, 50001 (2011).

\bibitem{spin1} S. A. Wolf, {\it et al.} Science \textbf{294},
1488 (2001).

\bibitem{spin2} D. E. Nikonov, G. I. Bourianoff, and P. A. Gargini,
J. Supercond. Novel Magn. \textbf{19}, 497-513 (2006).

\bibitem{metal} B. Behin-Aein, D. Datta, S. Salahuddin, S. and S.
Datta, Nature Nanotech. \textbf{6}, 266 (2010).

\bibitem{cite5} C. Joachim, J. K. Gimzewski, and H. Tang, Phys. Rev. B
\textbf{58}, 16407 (1998).

\bibitem{dsoi} G. Dresselhaus, Phys. Rev. \textbf{100}, 580 (1955).

\bibitem{rsoi} Y. A. Bychkov and E. I. Rashba, JETP Lett. \textbf{39},
78 (1984).

\bibitem{gate1} Z. Scher\"ubl, G. F\"ul\"op, M. H. Madsen, J. Nyga\r{r}d,
and S. Csonka, Phys. Rev. B. \textbf{94}, 035444 (2016).

\bibitem{gate2} T. W. Chen, C. M. Huang, and G. Y. Guo, Phys. Rev. B
\textbf{73}, 235309 (2006).

\bibitem{repro} A. Ney, C. Pampuch, R. Koch, and K. H. Ploog,
Nature \textbf{425}, 485 (2003).

\bibitem{ham1} J. S. Sheng and K. Chang, Phys. Rev. B \textbf{74}, 
235315 (2006).

\bibitem{ham2} C. P. Moca and D. C. Marinescu, J. Phys.: Condens. Matter
\textbf{18}, 127 (2006).

\bibitem{ham3} S. K. Maiti, J. Appl. Phys. \textbf{110}, 064306 (2011).

\bibitem{ham4} M. Patra and S. K. Maiti, Eur. Phys. J. B \textbf{89}, 
88 (2016).

\bibitem{ph} S. K. Maiti, S. Saha, and S. N. Karmakar, Eur. Phys. J.
B \textbf{79}, 209 (2011).

\bibitem{datta} Datta, S. Electronic transport in mesoscopic systems
(Cambridge University Press, Cambridge, 1995).

\bibitem{car} C. Caroli, R. Combescot, P. Nozieres, and D. Saint-James, 
J. Phys C: Solid State Phys. \textbf{4}, 916 (1971).

\bibitem{fl} D. S. Fisher and P. A. Lee, Phys. Rev. B \textbf{23}, 6851 (1981).

\bibitem{chnew} M. Wang and K. Chang, Phys. Rev. B \textbf{77}, 125330 (2008).

\bibitem{chnew1} W. Yang and K. Chang, Phys. Rev. B \textbf{73}, 045303 
(2008).

\bibitem{spl1} A. A. Kislev and K. W. Kim, J. App. Phys. \textbf{94}, 4001
(2003).

\bibitem{spl2} I. A. Shelykh, N. G. Galkin, and N. T. Bagraev,
Phys. Rev. B \textbf{72}, 235316 (2005).

\bibitem{spl3} P. F\"{o}ldi, O. K\'{a}lm\'{a}n, M. G. Benedict, and
F. M. Peeters, Phys. Rev. B \textbf{73}, 155325 (2006).

\bibitem{spl4} M. Dey, S. K. Maiti, S. Sil, and S. N. Karmakar, 
J. Appl. Phys. \textbf{114}, 164318 (2013).

\bibitem{spl5} S. K. Maiti, Phys. Lett. A \textbf{379}, 361 (2015). 

\bibitem{gate3} J. Chen, {\it et al.}, Phys. Rev. Lett. \textbf{105},
176602 (2010).

\bibitem{gate4} Y.-J. Yu, {\it et al.}, Nano Letters. \textbf{9},
3430 (2009).

\bibitem{qnano} P. F\"{o}ldi, O. K\'{a}lm\'{a}n, M. G. Benedict, and
F. M. Peeters, Nano Lett. \textbf{8}, 2556 (2008).

\bibitem{qnano1} O. K\'{a}lm\'{a}n, P. F\"{o}ldi, M. G. Benedict, and
F. M. Peeters, Physica E \textbf{40}, 567 (2008).

\bibitem{stt1} N. Locatelli, V. Cros, and J. Grollier, Nat. Mater.
\textbf{13}, 11 (2014).

\bibitem{stt2} Memory with a spin. Editorial. Nature Naotech. \textbf{10},
185 (2015).

\bibitem{stt3} M. Patra and S. K. Maiti, Europhys. Lett. \textbf{121}, 
38004 (2018).

\bibitem{stt4} D. C. Ralph and M. D. Stiles, M. D., J. Magn. Magn.
Mater. \textbf{320}, 1190 (2008).

\end{thebibliography}
\end{document}